\documentclass{JHEP3}
\pdfoutput=1

\usepackage{amsmath}

\usepackage{feynmf}
\usepackage{graphicx}

\newcommand{\be}{\begin{equation}}
\newcommand{\ee}{\end{equation}}
\newcommand{\br}{\begin{eqnarray}}
\newcommand{\bea}{\begin{eqnarray}}
\newcommand{\eea}{\end{eqnarray}}
\newcommand{\er}{\end{eqnarray}}
\newcommand{\ba}{\begin{array}}
\newcommand{\ea}{\end{array}}
\newcommand{\bi}{\begin{itemize}}
\newcommand{\ei}{\end{itemize}}
\newcommand{\bn}{\begin{enumerate}}
\newcommand{\en}{\end{enumerate}}
\newcommand{\bc}{\begin{center}}
\newcommand{\ec}{\end{center}}

\newcommand{\Eq}[1]{Eq.~(\ref{#1})}
\newcommand{\rfn}[1]{(\ref{#1})}

\def\gappeq{\mathrel{\rlap {\raise.5ex\hbox{$>$}}
{\lower.5ex\hbox{$\sim$}}}}

\def\lappeq{\mathrel{\rlap{\raise.5ex\hbox{$<$}}
{\lower.5ex\hbox{$\sim$}}}}

 \def\nn{\nonumber}

 \def\M{\mathcal{M}}
 \def\a{\alpha}

 \def\g{\gamma}
 \def\G{\Gamma}
 \def\d{\delta}
 \def\D{\Delta}
 \def\m{\mu}
 \def\n{\nu}

 \def\l{\lambda}

 \def\s{\sigma}

 \def\({\left(}
 \def\){\right)}
 \def\[{\left[}
 \def\]{\right]}

 \def\ds#1{#1\kern-1ex\hbox{/}}
\def\sla{\raise.15ex\hbox{$/$}\kern-.57em}
\newcommand{\sdm}{S_{\text{DM}}}
\newcommand{\snl}{S_{\text{NL}}}
\newcommand{\snltw}{S_{\text{NL2}}}
\newcommand{\snlth}{S_{\text{NL3}}}

\newcommand{\gev}{\text{GeV}}
\newcommand{\tev}{\text{TeV}}

\newcommand{\hc}[1]{#1^{\dagger}} % Hermitian conjugate

\newcommand{\gsim}{\lower.7ex\hbox{$\;\stackrel{\textstyle>}{\sim}\;$}}
\newcommand{\lsim}{\lower.7ex\hbox{$\;\stackrel{\textstyle<}{\sim}\;$}}

\title{Long-lived charged Higgs at LHC as a probe of  scalar  Dark Matter}

\author{{\bf K. Huitu$^{a,b}$},  {\bf K. Kannike$^c$}, {\bf A. Racioppi$^c$},
and {\bf M. Raidal$^{a,c}$}\\
$^a$Department of Physics, P.O.Box 64, FIN-00014 University of Helsinki, Finland \\
$^b$Helsinki Institute of Physics, P.O.Box 64, FIN-00014 University of
Helsinki, Finland \\
$^c$National Institute of Chemical Physics and Biophysics, Ravala 10,
Tallinn 10143, Estonia}

\preprint{HIP-2010-16/TH}

\abstract{
We study inert charged Higgs boson $H^\pm$ production and decays at LHC experiments  in the context of
constrained scalar dark matter model (CSDMM). In the CSDMM the mass spectrum of the inert doublet and singlet scalars
is predicted from the GUT scale initial conditions via RGE evolution. We compute the cross sections of processes
$pp\to H^+H^-,\, H^\pm S_i^0,$ where $S_i^0$ are neutral scalar particles,  at the LHC experiments.
We show that for light $H^\pm$ the first process may receive a sizable contribution from the top quark mediated 1-loop  diagram with
Higgs boson in $s$-channel. In a significant fraction of the parameter space  $H^\pm$ are long-lived
because their decays to predominantly singlet scalar dark matter (DM) and next-to-lightest (NL) scalar,
$H^\pm\to S_{\text{DM, NL}} ff',$ are suppressed
by the small singlet-doublet mixing angle and by the moderate mass difference $ \Delta M=M_{H^+}-M_{\text{DM}} .$
The experimentally measurable displaced vertex in $H^\pm$ decays to leptons and/or jets and missing energy allows one
to discover the $H^+H^-$ signal over the huge $W^+W^-$ background.
If, however, $H^\pm$ are short-lived, the subsequent decays $S_{\text{NL}}\to S_{\text{DM}} f\bar f$
necessarily produce additional displaced vertices that allow to reconstruct the full $H^\pm$ decay chain.
We propose benchmark points for studies of
this scenario at the LHC.
}


\begin{document}

\section{Introduction}

The existence of cold dark matter (DM) of the Universe
is firmly established by cosmological observations \cite{Komatsu:2008hk}. Because the SM
does not contain a cold DM candidate, its existence is a clear signal of new physics beyond the SM.
However, the origin, nature and properties of the DM have  so far remained completely unknown.
The Tevatron and the LHC experiments aim to reveal the origin of electroweak symmetry breaking (EWSB) and to discover the DM particle directly.
In the standard model (SM) there is just one fundamental scalar doublet $H_1$ and the  EWSB occurs
spontaneously due to its explicitly negative mass parameter $\mu^2_1$ in the scalar potential. In the SM the Higgs boson,
the only scalar particle to be discovered, cannot be the DM candidate.

However, it is possible that the two issues are  related in models with an extended scalar sector.
The SM Higgs boson mass term $\mu_1^2 H_1^\dagger H_1,$
being superrenormalizable, may open a portal into a hypothetical hidden sector \cite{Patt:2006fw}.
Combining this idea with a scenario that cold DM of the Universe  consists of
a $Z_2$-odd SM singlet $S$~\cite{McDonald:1993ex,Burgess:2000yq,Barger:2007im,Barger:2008jx} and/or doublet  $H_2$~\cite{Deshpande:1977rw,Ma:2006km,Barbieri:2006dq,LopezHonorez:2006gr} scalars implies that
the SM Higgs boson opens a portal into DM. It is also possible that the new scalar DM sector actually
triggers the EWSB by driving  $\mu^2_1$ negative by some dynamical mechanism.
In order to formulate this interesting but purely phenomenological scenario in the form of DM theory
one needs  theoretical guidance from the underlying principles of new physics.

It was shown in \cite{Kadastik:2009dj,Kadastik:2009cu}  that the high energy theory for the low scale scalar DM models
can be non-SUSY $SO(10)$ Grand Unified Theory (GUT) \cite{Fritzsch:1974nn}.
Indeed,  the discrete $Z_2$ symmetry, which makes DM stable, could be
an unbroken remnant of some underlying $U(1)$ gauge subgroup \cite{Krauss:1988zc,Martin:1992mq,DeMontigny:1993gy}.
This argument is completely general and does not require the existence of additional symmetries such
as supersymmetry\footnote{In the context of minimal supersymmetric standard model (MSSM) R-parity \cite{Farrar:1978xj} is imposed
by hand to prevent phenomenological disasters such as a rapid proton decay.
In MSSM the R-parity is equivalent to the matter-parity \cite{Bento:1987mu,Ibanez:1991hv}  that is imposed at superfield level.}.
If the GUT gauge group is $SO(10)$,
the argument of  \cite{Krauss:1988zc,Martin:1992mq,DeMontigny:1993gy} implies \cite{Kadastik:2009dj,Kadastik:2009cu}
that non-supersymmetric DM should most naturally be embedded into a scalar
representation ${\bf 16}$ because this is the only small representation that is odd under the generated discrete gauge symmetry -- the matter-parity
\bea
P_M=(-1)^{3(B-L)}.
\label{PM}
\eea
In this framework the generation of matter-parity $P_M$ is directly related to the  breaking of gauged $B-L,$  implying that the dark sector
actually consists of $P_M$-odd scalar relatives of the SM fermions\footnote{In the context of supersymmetry the scalar particles with
the same quantum numbers are called squarks and sleptons. This scenario suggests generally that what we call ``matter" must consist of $P_M$-odd particles. }.
In this scenario the origin and stability of DM, the non-vanishing neutrino masses via the seesaw mechanism \cite{Minkowski:1977sc,Yanagida:1979,GellMann:1980vs,Glashow:1979nm,Mohapatra:1979ia} and
the baryon asymmetry of the Universe via leptogenesis \cite{Fukugita:1986hr} all originate from
the same source -- the breaking of  $SO(10)$ gauge symmetry. In addition, the EWSB may occur dynamically due to the Higgs boson
interactions with the dark sector scalars~\cite{Kadastik:2009cu,Hambye:2007vf,Kadastik:2009ca}.

The inert charged Higgs boson production and decays at the LHC experiments have been previously studied  in three papers
\cite{Cao:2007rm,Dolle:2009ft,Miao:2010rg}.  Working  in the context of the inert doublet model \cite{Deshpande:1977rw,Ma:2006km,Barbieri:2006dq,LopezHonorez:2006gr}, those papers conclude that
it is impossible to discover the production processes $pp\to H^+H^-, \,H^\pm S$ followed by the decays
$H^\pm \to S_\text{DM} \ell^\pm \nu,$ where $S_\text{DM}$ is the DM scalar, over the huge  $W^+W^-$ background.
However, as explained above, the  inert doublet model represents just one particular corner of parameter space of the
general $P_M$-odd  scalar  DM scenario in which the DM is (predominantly) doublet.

From the fundamental physics point of view a better motivated scalar particle spectrum is obtained from the GUT scale initial
 parameters by their renormalization group (RG) evolution down to the low scale \cite{Kadastik:2009cu}.
 This procedure is analogous to obtaining the low scale sparticle spectrum in the constrained MSSM.
In our scenario the constrained scalar DM model (CSDMM) predicts  that in the majority of parameter space the DM  scalar
is predominantly singlet and that $H^\pm$ and $S_\text{DM}$ are relatively close in mass.
Therefore the decays $H^\pm \to S_\text{DM,NL} f f',$ where  the  next-to-lightest neutral scalar $S_\text{NL}$ is
almost degenerate with $S_\text{DM}$ and $ff'$ are the SM quarks and leptons,   may be suppressed by two factors:
$(i)$ by the small singlet-doublet mixing angle; $(ii)$ by the small mass difference $ \Delta M=M_{H^+}-M_{\text{DM}} .$
Thus the inert  $H^\pm$ may be long-lived, travel a macroscopic distance inside the tracker of a LHC experiment,
 and decay far from the interaction point into charged lepton or jets and missing $E_T$.
 The experimental signature of the  displaced vertices in $H^\pm$ decays are theoretically free
 from the SM background and enable to discover $H^\pm$ at the LHC.

In this work we study charged Higgs boson phenomenology of the constrained scalar DM model.
Because $H^+$ does not have  Yukawa couplings, we traditionally call it inert charged Higgs boson
even though it does have gauge couplings.
There are several differences between the  phenomenology of the charged Higgs boson of the
two Higgs doublet models (2HDM)  and the phenomenology of the inert charged Higgs boson.
First of all, the inert charged Higgs boson is coupled only to bosons (scalars or gauge vectors),
and its interactions are determined by  the matter-parity conservation.
Moreover, it is the charged component of only the $P_M$-odd
doublet and  does  not mix with the charged component
of the SM Higgs doublet. As a consequence, there are no free parameters such as $\tan\beta .$
These features will affect the inert charged Higgs production and decays to be studied in this paper.

We first review the basics of the constrained scalar DM model \cite{Kadastik:2009cu}.
 After that we study the production and the decays of the inert charged Higgs
boson at the LHC experiments. First we show that for light charged Higgs scalar the 1-loop top quark mediated production process with
the Higgs boson in the $s$-channel may be of the same order of magnitude as the tree level processes considered
in papers \cite{Cao:2007rm,Dolle:2009ft,Miao:2010rg}. This is an important new result of our paper
which agrees with the similar result obtained for the production of a pair of neutral inert scalars \cite{Kadastik:2009gx}.
After that we show that the constrained scalar DM model may imply a long lifetime of $H^+$ which, in an appreciable
 fraction of the parameter space, may imply observable displaced vertices at the LHC. This is a background-free experimental signature
of the inert charged Higgs boson. Further we argue that if $H^\pm$ is short-lived, the decay chain $H^\pm \to S_\text{NL} f f'$
followed by $S_{\text{NL}}\to S_{\text{DM}} f\bar f$ will necessarily produce a displaced vertex in the latter decay
that  allows to reduce the background.
Throughout of this paper we consider two scenarios of the constrained scalar DM models,
one without requiring radiative EWSB and another with the requirement of radiative EWSB. The latter is considerably more constrained
because the negative SM Higgs boson mass parameter $\mu^2_1<0$ is obtained via the RG effects of the dark scalar sector \cite{Kadastik:2009cu}.
Finally we propose three benchmark scenarios which allow one to discover the long-lived $H^\pm$ at the LHC both in the case of
radiative and explicit EWSB.

\section{The constrained scalar Dark Matter model (CSDMM)}

The CSDMM is obtained from the minimal non-supersymmetric matter-parity-odd scalar $SO(10)$ model
by decoupling of all new  $SU(2)_L\times U(1)_Y$ scalar multiplets that do not contain the DM candidates (that this is possible remains to be shown in a detailed $SO(10)$ model).
This implies that the scalar sector of CSDMM consists of the Higgs boson $H_1$, the inert doublet $H_2$ and the
singlet $S.$ The minimal $P_M$-odd scalar $SO(10)$ GUT  scenario\footnote{An alternative possibility to introduce non-supersymmetric DM
using the general idea of non-supersymmetric matter-parity $P_M$ proposed in  \cite{Kadastik:2009dj}
 is  to introduce $P_M$-even fermion multiplets of $SO(10)$ \cite{Frigerio:2009wf}.}
contains the SM Higgs boson in a scalar  representation ${\bf 10}$ and  the DM candidates in a scalar representation ${\bf 16}$.
Thus below the  $M_\text{G}$ and above the EWSB scale
the model is described by the $H_{1} \to H_{1}$, $S \to -S,$ $ H_{2} \to -H_{2}$
invariant scalar potential
\begin{equation}
\begin{split}
V = &\, \mu_{1}^{2} \hc{H_{1}} H_{1} + \lambda_{1} (\hc{H_{1}} H_{1})^{2}
+ \mu_{2}^{2} \hc{H_{2}} H_{2} + \lambda_{2} (\hc{H_{2}} H_{2})^{2} \\
&+ \mu_{S}^{2} \hc{S} S + \frac{\mu_{S}^{\prime 2}}{2} \left[ S^{2} + (\hc{S})^{2} \right] + \lambda_{S} (\hc{S} S)^{2}
 \label{V}\\
& + \frac{ \lambda'_{S} }{2} \left[ S^{4} + (\hc{S})^{4} \right]
 + \frac{ \lambda''_{S} }{2} (\hc{S} S) \left[ S^{2} + (\hc{S})^{2} \right] \\
&+ \lambda_{S1}( \hc{S} S) (\hc{H_{1}} H_{1}) + \lambda_{S2} (\hc{S} S) (\hc{H_{2}} H_{2}) \\
&+ \frac{ \lambda'_{S1} }{2} (\hc{H_{1}} H_{1}) \left[ S^{2} + (\hc{S})^{2} \right]
+ \frac{ \lambda'_{S2} }{2} (\hc{H_{2}} H_{2}) \left[ S^{2} + (\hc{S})^{2} \right]
 \\
&+ \lambda_{3} (\hc{H_{1}} H_{1}) (\hc{H_{2}} H_{2}) + \lambda_{4} (\hc{H_{1}} H_{2}) (\hc{H_{2}} H_{1}) \\
&+ \frac{\lambda_{5}}{2} \left[(\hc{H_{1}} H_{2})^{2} + (\hc{H_{2}} H_{1})^{2} \right] \\
&+ \frac{\mu_{S H}}{2} \left[\hc{S} \hc{H_{1}} H_{2} + {S} {H_{1}} \hc{H_{2}}    \right]
+ \frac{\mu'_{S H}}{2} \left[S \hc{H_{1}} H_{2} + \hc{S} {H_{1}} \hc{H_{2}}  \right],
\end{split}
\end{equation}
together with the GUT scale boundary conditions
\bea
&\mu_1^2(M_{\text{G}})>0,\; \mu_2^2(M_{\text{G}})=\mu_S^2(M_{\text{G}}) >0, &
\label{bc1}\\
&\lambda_2(M_{\text{G}})=\lambda_S(M_{\text{G}})=\lambda_{S2}(M_{\text{G}}),\; \lambda_3(M_{\text{G}})=\lambda_{S1}(M_{\text{G}}), \nonumber&
\eea
and
\begin{equation}
\begin{split}
{\mu}_S^{\prime 2}, \; {\mu}_{SH}^{ 2} &\lsim {\cal O}\left( \frac{M_\text{G}}{M_{\text{P}}}\right)^n \mu^2_{1,2},
\\
 \lambda_{5} ,\; \lambda'_{S1} ,\; \lambda'_{S2} ,\; \lambda''_{S} &\lsim {\cal O}\left( \frac{M_\text{G}}{M_{\text{P}}}\right)^n \lambda_{1,2,3,4}.
\end{split}
\label{bc2}
\end{equation}
While the parameters in \Eq{bc1} are allowed by
$SO(10),$ the ones in \Eq{bc2} can be generated only after $SO(10)$ breaking by  operators
suppressed by $n$ powers of the Planck scale $M_{\text{P}}.$
Because the low scale particle mass spectrum is obtained from a small number of parameters at  $M_\text{G}$ via RGE running  \cite{Kadastik:2009cu},
this scenario is called constrained scalar DM model in a direct analogy with
the constrained MSSM\footnote{To achieve unification of the gauge couplings, threshold effects of 20\% are needed for $\alpha_1$. This is of the
same order as in the SM. The extra fields near the GUT scale needed to make this possible can in general also influence the low energy particle spectrum.
We assume here that this in influence is insignificant.}.

The charged Higgs boson mass coming from \Eq{V} is  given by
\bea
m_{H^{+}}^{2} = \mu_{2}^{2} + \lambda_{3} v^{2}/2,
\label{mch}
\eea
and the  neutral $P_M$-odd scalar masses $m_{1,2}^{2}$ are obtained by diagonalization of the mass matrix
\begin{equation}
m^2=\begin{pmatrix}
\mu_{2}^{2}+ (\lambda_{3} + \lambda_{4} + \lambda_{5}) v^{2}/2
& (\mu_{SH} + \mu'_{SH}) v/(2 \sqrt{2}) \\
 (\mu_{SH} + \mu'_{SH}) v/(2 \sqrt{2})
 & \mu_{S}^{2} + \mu_{S}^{\prime 2} + (\lambda_{S1} + \lambda'_{S1} ) v^2/2
 \end{pmatrix}.
 \label{mn}
\end{equation}
The pseudo-scalar masses $m_{3,4}^{2}$ are obtained from $m_{1,2}^{2}$
by replacing $\lambda_5 \to -\lambda_5,$
$\lambda'_{S1} \to -\lambda'_{S1},$  $ \mu_{S}^{\prime 2} \to  -\mu_{S}^{\prime 2}.$
 For clarity we  denote the lightest neutral scalar by $S_{\text{DM}}$ and the next-to-lightest neutral scalar by $S_{\text{NL}}.$

We note that the mass degeneracy of $S_{\text{DM}}$ and $S_{\text{NL}}$  is a generic {\it prediction} of the scenario and follows from
the underlying $SO(10)$ gauge symmetry via \Eq{bc2}. This degeneracy has several phenomenological implications
which allow one to discriminate this scenario from other DM models. For example, it implies a long lifetime for
 $S_{\text{NL}}$ which provides clear experimental signature of displaced vertex in the decays $S_{\text{NL}}\to S_{\text {DM}} \ell^+ \ell^-$
 at the LHC \cite{Kadastik:2009gx}.
In the context of present work  the decays of $S_{\text{NL}}$ occur in the chain of $H^\pm$ decays and allow one to
distinguish $H^\pm$ over the SM background.

 At $M_{\text{G}}$ the SM gauge symmetry may not be spontaneously broken, $\mu_1^2(M_{\text{G}})>0.$
 To obtain successful EWSB at low energies the parameter $\mu_1^2(M_Z)$ can become negative
 either by the RG evolution \cite{Kadastik:2009cu} or via the Coleman-Weinberg-like  \cite{Coleman:1973jx} EWSB mechanism
\cite{Hambye:2007vf,Kadastik:2009ca}. Thus in the constrained scalar DM model the EWSB may occur dynamically due to the
 existence of dark scalar couplings to the SM Higgs boson. In the following we study two scenarios of the CSDMM, one with explicit
 EWSB as in the SM, and one with radiative EWSB due to DM RG effects in the scalar sector.

\section{Inert charged Higgs boson phenomenology at LHC}

\FIGURE[t!]{ \centering
\includegraphics[scale=0.45]{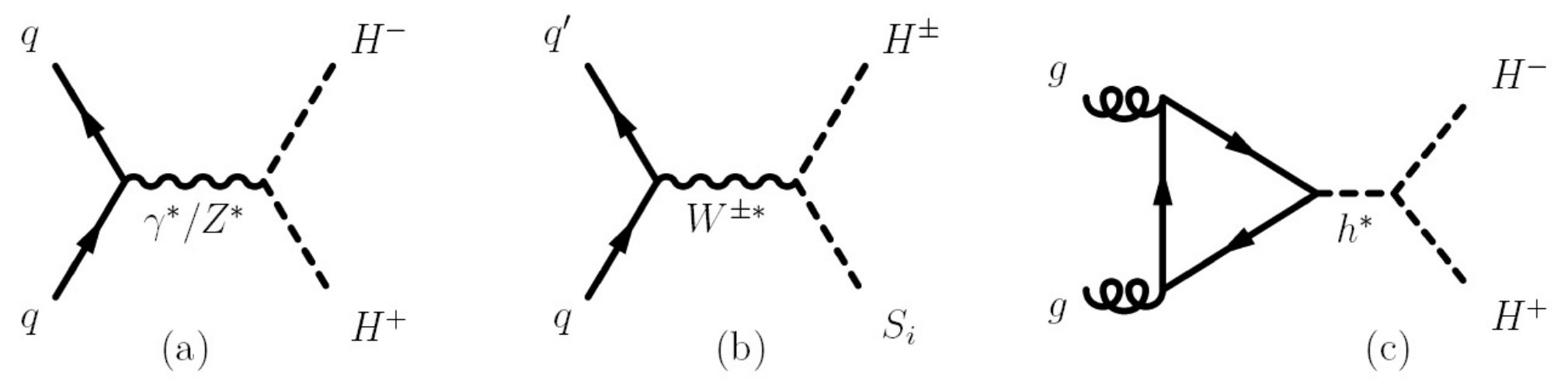}
\caption{Feynman diagrams for inert charged Higgs  boson production at the LHC.}
\label{CHprod}
}

\begin{figure}[t!]
\centering
\includegraphics[scale=0.3]{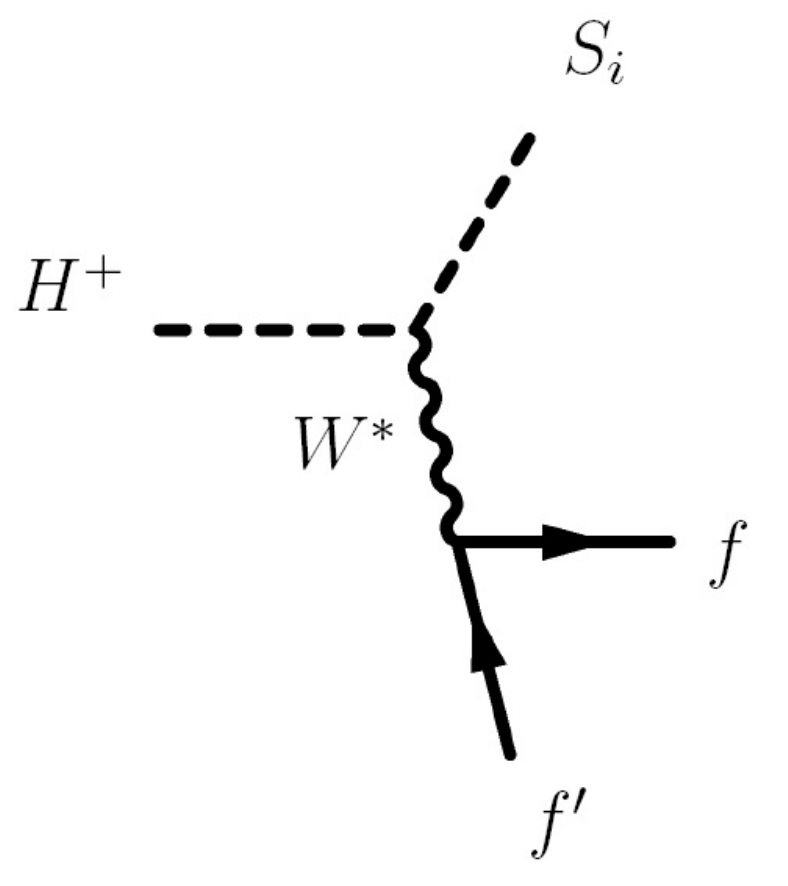}
\caption{Feynman diagram for the inert $H^+$ decays into  dark scalars $S_i$ and two fermions.}
\label{Hpdecay}
\end{figure}

Compared to the 2HDM models, inert $H^\pm$ production lacks the primary parton level
process $b \bar b \to H^+ H^-$ with a top quark in the $t$-channel and the
secondary production processes through the decays $t \to H^+  b$. The only available production processes are
depicted in Fig.~\ref{CHprod}.
Because of the matter-parity conservation,
the inert charged Higgs can only decay into an odd (usually one)
number of dark scalars plus SM particles, see Fig.~\ref{Hpdecay}.
Thus the 2HDM decays like $H^+ \to W^+
H^0$ or $H^+ \to t \bar b$ cannot take place.
We are going to show that those features allow to discover matter-parity-odd charged Higgs
boson at the LHC.

\subsection{Direct production}

The main parton level production processes for the inert charged scalars at the LHC are (see Fig.~\ref{CHprod}):
\begin{subequations}
\be
  q \bar{q} \rightarrow \g^*/Z^* \rightarrow  H^{+} H^{-}, \label{eq:qqHH}
\ee
\be
  q \bar{q}' \rightarrow W^\pm \rightarrow S_{i} H^\pm, \label{eq:qqSH}
\ee
\be
  g g \rightarrow h^* \rightarrow  H^{+} H^{-}, \label{eq:ggHH}
\ee
\end{subequations}
where $S_i$ stands for  the new $P_M$-odd neutral scalars and $h$ is   the SM Higgs boson.
The processes \rfn{eq:qqHH} and \rfn{eq:qqSH} have previously been studied in Refs. \cite{Cao:2007rm,Dolle:2009ft}.
We present details of calculating cross sections of those processes for completeness.
Studies of the process \rfn{eq:ggHH}, which may give the dominant contribution to the production cross section
for very light charged Higgs at the resonance of the SM Higgs boson, is a new result of this paper.
The process \rfn{eq:ggHH} is proportional to the single parameter $\lambda_3$ in \Eq{V} and may allow one to measure
that parameter at the LHC experiments. We note that
the process $g g \to \gamma^*/Z^*  \to H^+ H^-$
vanishes identically because of the assumed CP invariance (see for instance \cite{Krause:1997rc}).

We begin with discussing the process (\ref{eq:qqHH}).
Its parton level cross section is given by
 \bea \s_{q \bar{q} \to H^+ H^-}&&=
-\frac{e^4 }{2304 \pi  c_W^4 s_W^4}
 \frac{ ( \hat{s}( \hat{s}-4 M_{H^\pm}^2 ) )^{3/2}}{ \hat{s}^{9/2}  \sqrt{\hat {s}-4m_q^2} ( M_{Z}^2-\hat{s} )^2} \nn \\
 &&\times
 \[ a_q^2 \right. \hat{s}^2 ( c_W^2-s_W^2 )^2 ( 4 m_q^2-\hat{s} )  -
   ( 2 m_q^2+\hat{s} ) \(4 q_q c_W^2 s_W^2 ( \hat{s}-M_{Z}^2 )^{2} \right.
\nn \\
  &&\left.+v_q  \hat{s}  ( c_W^2-s_W^2 ) )^2 \], \label{6a}
\eea
where $\hat s$ is the usual kinematical variable, and $M_{H^\pm}$, $M_Z$ and $m_q$  are the $H^\pm$, $Z$ and quark masses, respectively.
The  cross section in $pp$ collisions is obtained by convoluting over parton structure functions $q(x)$ as
\be
 \s_{pp \to  H^+ H^-}^{q \bar q}= \int d x_a d x_b [q_a (x_a) \bar q_b (x_b)+q_b (x_b)\bar q_a (x_a) ] \s_{q \bar{q} \to H^+ H^-}.
\ee

For the process (\ref{eq:qqSH})
the corresponding parton level cross section is
\bea
&&\!\!\!\!\!\!\!\!\!\!\!\!\s_{q \bar q' \to S_i H^+}=\s_{q' \bar q \to S_i H^-}=\frac{\eta_i^2}{4608 \pi } \( \frac{e}{s_W} \)^4
   \label{6b}  \\
  &&\!\!\!\!\!\!\!\!\!\!\!\! \times
    \frac{ \( M_{H^\pm}^4-2 M_{H^\pm}^2 \( M_{S_i}^2+{\hat s} \)+
            \( M_{S_i}^2-{\hat s} \)^2 \)^{3/2} \(2 {\hat s}^2- {\hat s} \( m_q^2+m_{q'}^2 \)-\( m_q^2-m_{q'}^2 \)^2\)}{
     {\hat s}^3  \( M_W^2-{\hat s} \)^2 \sqrt{-2 {\hat s} \( m_q^2+m_{q'}^2 \)+\( m_q^2-m_{q'}^2 \)^2+{\hat s}^2}}, \nn
\eea
where $q (q')$ means up(down)-type quark and $M_W$ is the $W$ boson mass.
Notice that the cross section \rfn{6b} depends on the nature of the final state neutral scalar $S_i.$
If the outgoing scalar is singlet-like one has $\eta_i=s,$ where $s$ is the sine of the small singlet-doublet mixing angle, and
the cross section \rfn{6b} is very much suppressed.
If, however,  the outgoing scalar is doublet-like, $\eta_i=c$ is of order unity.
Since the process with an outgoing $H^+$ is the conjugate of the process with an outgoing $H^-$, the corresponding parton level
cross sections are equal. However, the observable cross sections in $pp$ collisions at the LHC
are different because of the different parton structure functions for  up and down type quarks,
\bea
 \s_{pp \to S_i H^+}^{q \bar q'} &=& \int d x_a d x_b \[ q_a (x_a) \bar q'_b (x_b) + q_b (x_b) \bar q'_a (x_a) \] \s_{q \bar q' \to S_i H^+}, \\
 \s_{pp \to S_i H^-}^{q' \bar q} &=& \int d x_a d x_b \[ q'_b (x_b) \bar q_a (x_a) + q'_a (x_a) \bar q_b (x_b) \] \s_{q' \bar q \to S_i H^-}.
\eea

Finally we study the process (\ref{eq:ggHH}).
The Feynman amplitude of that process is given by
\be
 |\M_{gg \to H^+ H^-}|^2 =L^2 \( \frac{\lambda_{3}v}{\hat s-M_h^2} \)^2,
 \label{lam3ref}
\ee
where  $M_h$ is the Higgs boson mass, $\lambda_{3}v$ is the trilinear scalar self-coupling,
 and the loop factor $L$ is given by \cite{Gunion:1989we}
\be
 L^2 =\frac{\a_S}{8 \pi^2 v} \left|\sum_q F_q \right|^2 \quad , \qquad \qquad
 F_q = - (2 m_q)^2 \[ 1+ (1-\tau) f(\tau)\],
\ee
where
\vskip -0.9cm
\be
 f(\tau)=  \left\{ \begin{array}{cc}
                 \[ \sin^{-1} \( \sqrt{1/\tau} \) \]^2 & \tau \geq 1 \\
                 -\frac{1}{4} \[ \log \( \frac{ 1+\sqrt{1-\tau} }{ 1+\sqrt{1-\tau} } \) - i \pi  \]^2   & \tau < 1
                 \end{array}
          \right.
\quad , \qquad \qquad \tau = \frac{(2 m_q)^2}{\hat s}.
\ee
\vskip -0.1cm
The corresponding parton level cross section reads
\be
  \s_{gg \to H^+ H^-}=\frac{\a_S^2 \( \lambda_{3}v \)^2}{32768 \, \pi ^3}
   \frac{\left|\sum_q F_q \right|^2 \sqrt{\( -2 M_{H^\pm}^2+{\hat s} \)^2-4 M_{H^\pm}^4}}{ {\hat s}^2 v^2 \( {\hat s}-M_h^2 \)^2},
\ee
while the integrated cross section is found via
\be
 \s_{pp \to H^+ H^-}^{g g}= \int d x_a d x_b [g_a (x_a) g_b (x_b)+g_b (x_b) g_a (x_a) ] \s_{gg \to H^+ H^-},
\ee
where $g(x)$ is the gluon density function.

We computed the cross sections  of the processes  \rfn{eq:qqHH},  \rfn{eq:qqSH} and \rfn{eq:ggHH}  in $pp$ collisions at the LHC
by convoluting over the parton distribution functions of Ref. \cite{Alekhin:2006zm}.
We scan over the parameter space of the model and select out the parameters
that imply the observed  amount of DM, $0.094<\Omega_\text{DM}<0.129,$ in thermal freeze-out at early Universe.
The DM abundance is  calculated with MicrOMEGAs package \cite{Belanger:2006is,Belanger:2008sj}.
As already explained, we consider separately the scenarios with and without radiative EWSB mechanism.

\FIGURE[t]{
\centering
\includegraphics[scale=0.35]{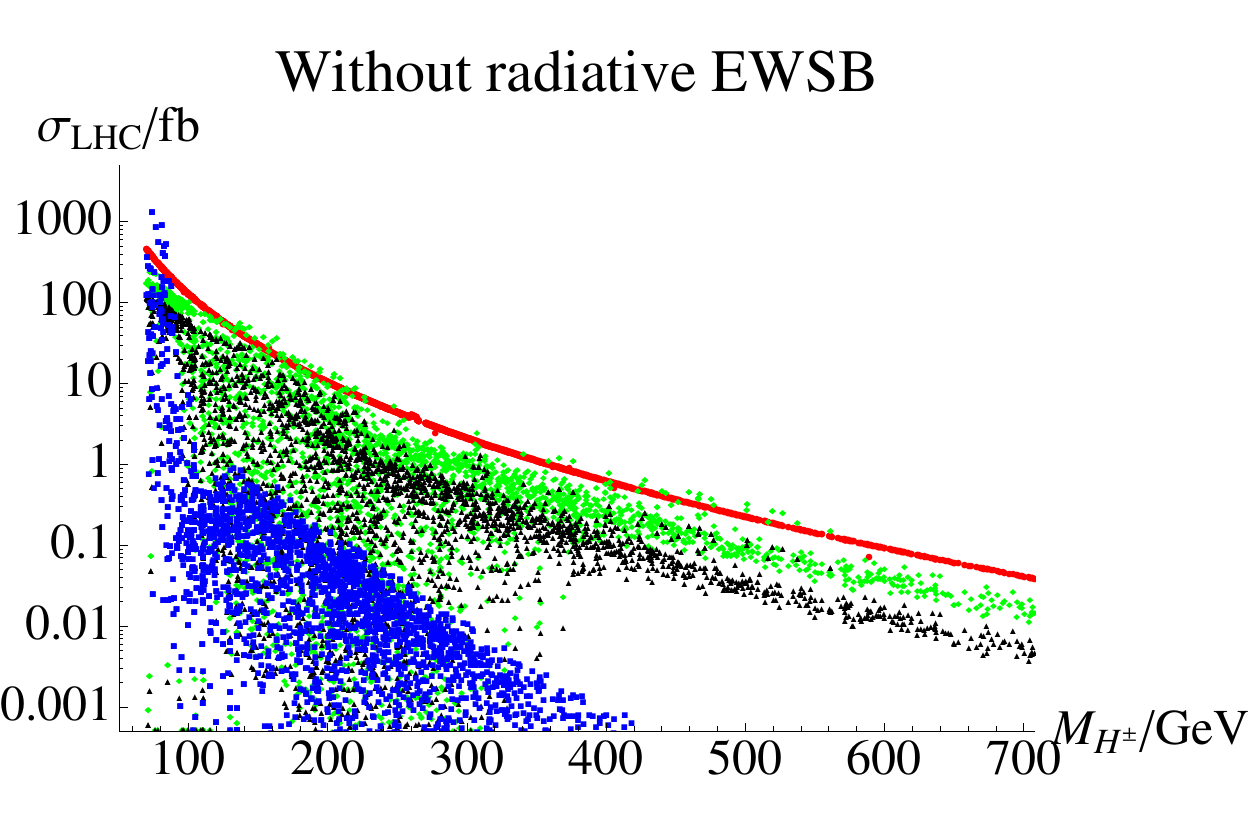}
\includegraphics[scale=0.35]{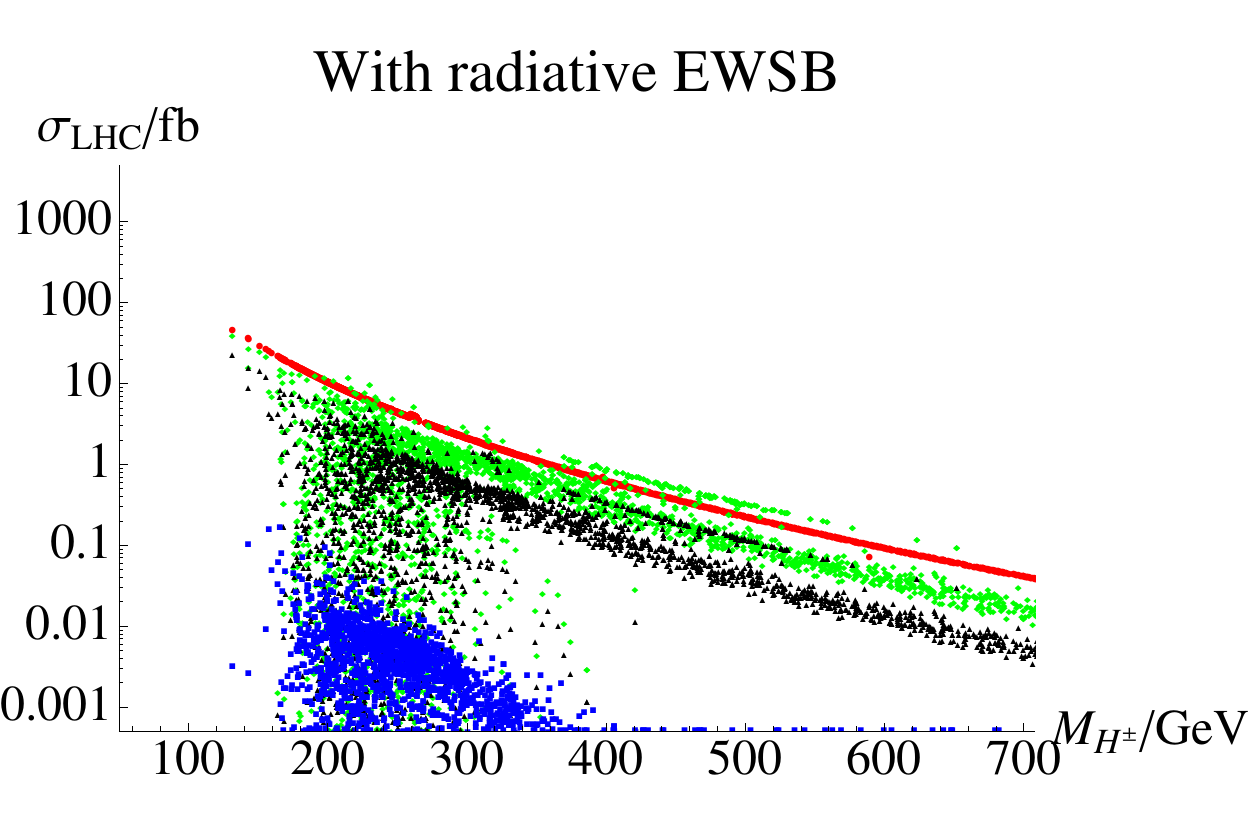}
\vspace{-0.5cm}
\caption{Cross-sections  for
 $pp \to H^{+} H^{-}$ via $q \bar q$ (red), $pp \to H^{+} H^{-}$ via $gg$ (blue),
$pp \to S_{\text{DM},\text{NL}} H^+$ (green) and $pp \to S_{\text{DM},\text{NL}} H^-$ (black) at the LHC for $\sqrt{s} = 14~\tev$
without (left panel) and with (right panel) radiative EWSB mechanism.
}
\label{fig:cshp:lhc}
}

Fig.~\ref{fig:cshp:lhc} shows the scatter plots of the $H^+H^-,$  $S_{\text{DM},\text{NL}} H^+$ and $S_{\text{DM},\text{NL}} H^-$
production cross sections in $pp$ collisions
at the LHC  for the collision energy $\sqrt{s} = 14~\tev$ as a function of charged Higgs mass.
The colour code is explained in the caption. Because $S_{\text{DM}}$ and $S_{\text{NL}}$ are almost degenerate, their production cross sections are almost equal.
The results of a general scan are presented in the left panel of Fig. \ref{fig:cshp:lhc}. In the right panel of the same figure successful EWSB is required
to  occur due to the RGE effects of the SM Higgs boson couplings to the DM sector.
One sees that no light charged Higgs ($M_{H^\pm}\lesssim 150~\gev$) is permitted in the latter case.
For very light $H^\pm$ the loop level gluon-gluon production process dominates over the tree level Drell-Yan processes only in the general
case when EWSB occurs explicitly.
For those points the pair production cross section is enhanced by the SM Higgs boson resonance.
In that case the parameter $\lambda_3$ can be directly derived from the cross section measurement, see \Eq{lam3ref}.
If one  requires radiative   EWSB from the DM couplings, the dark scalar mass scale is higher and this process is  suppressed.

Fig. \ref{fig:cshp:lhc} demonstrates that for heavy $H^\pm$ the sub-process \rfn{eq:qqHH} always dominates.
However, the cross section of this process is fully determined by the gauge couplings and depends only on the mass
of charged Higgs boson. Therefore one can reliably calculate the cross section of the sub-process \rfn{eq:qqHH}
and to split the experimentally measured $H^+H^-$ pair production cross section between the two dominant
contributions \rfn{eq:qqHH} and \rfn{eq:ggHH}.

\subsection{Long-lived $H^\pm$ in the constrained scalar DM scenario}

With the integrated luminosity 100~fb$^{-1}$ and low $H^\pm$ mass the LHC can produce thousands of  $H^+H^-$ pairs.
In this scenario the mass difference between the charged Higgs and the dark matter scalar turned out to be less than the $W$ mass and
the decays with on-shell $W$ or $Z$ in the final state are kinematically forbidden.
At the leading order the only kinematically allowed decays are (see Fig.~\ref{Hpdecay})
\be
 H^\pm \to S_i f \bar f',
 \label{decay}
\ee
where $S_i$ stands for any of the four neutral $P_M$-odd  scalars ($\sdm$, $\snl$, $\snltw$, $\snlth$) and $f,f'$ denote the SM leptons or quarks.
In most of the cases the only kinematically allowed decays of $H^\pm$ are to the lightest states $S_{\text{DM}}$ and $S_{\text{NL}}.$
Because those states are almost degenerate in mass due to the GUT gauge symmetry, see \Eq{bc2}, the $H^\pm$ \mbox{branching ratios}
to those states are practically equal. Thus, if $S_i\equiv S_{\text{DM}}$ in \rfn{decay},
the resulting experimental signatures of the $H^+H^-$ pair production include  $\ell^+\ell^-$, $jjjj$ or $\ell^\pm jj$  final states plus missing $E_T.$
Unfortunately those experimental signatures cannot be seen over the huge $W^+W^-$ background \cite{Cao:2007rm,Dolle:2009ft}  unless some new distinctive  feature occurs which allows to suppress the background. We claim in this work that the distinctive feature
might be macroscopically long $H^\pm$ lifetime.  However, if $S_i\equiv S_{\text{NL}}$ in \rfn{decay} that happens with almost equal probability,
the above described experimental signatures are going to be supplemented by the decays $S_{\text{NL}}\to S_{\text{DM}} f\bar f$
which necessarily must have a displaced $\ell^+\ell^-$ or $jj$ vertex due to the $S_{\text{DM,NL}}$ mass degeneracy \cite{Kadastik:2009gx}.
In the latter case the experimental signature of the $H^+H^-$ pair production includes two {\it additional}  displaced
$\ell^+\ell^-$ or $jj$ vertices.

Unlike in the inert doublet model \cite{Cao:2007rm,Dolle:2009ft},  in the  constrained scalar DM model the lightest dark
scalar is predicted to be dominantly singlet by RG evolution of the model parameters.
As already mentioned, the charged Higgs and the DM masses, given by \Eq{mch} and \Eq{mn}, respectively,
turned out to be close to each other in a wide range of the parameter space.
Therefore the $H^\pm$ decays  are  suppressed by two factors  $(i)$ by the sine of
singlet doublet mixing angle, $\eta=s$;  $(ii)$ by the (possibly) small mass difference $ \Delta M=M_{H^+}-M_{S_i} > m_f+m_{f'}.$
We also stress that if, instead, $H^\pm$ decays to the heavier states $\snltw$, $\snlth$ are kinematically allowed (this happens when $\l_4<0$),
the decay rate is proportional to $c^2$ instead of $s^{2}$.  This happens because the heavier states are usually doublet-like,
thus $H^\pm$ decay fast and there is no displaced vertex.
In order to clarify the discussion  we quantify the parameter space in which this may happen for the radiative EWSB case (the non-radiative case is analogous).
Considering the  $H^\pm$ masses up to 700~GeV, around 12\% of the randomly generated points that pass all the experimental constraints as described above have
$\Delta M<10$ GeV and only around 1\% have, at the same time, $H^\pm$ decays to  $\snltw$ and $\snlth$ kinematically allowed.
Thus for about 11\% of all the randomly generated points there is an approximate mass degeneracy together with decay channels
only to  $\sdm$ and $\snl$. Moreover the number of points that also show a tiny mixing angle, which means a displaced vertex ($\ell \gtrsim 0.1$ mm), is
about 7\% of all the randomly generated points.

However, if we restrict ourselves in the low mass region $M_{H^\pm} < 300$ GeV that is relevant for LHC experiments, this happens in 16\% of the parameter space.
We stress that the small mass splitting $\Delta M$ is not protected by any symmetry -- it is obtained by RGE analyses from the initial conditions
as described in the previous section, and is  \emph{accidental and model dependent} in nature.
However, our result shows that in the CSDMM for the mass range testable at LHC this happens almost for one sixth of the randomly generated points, which is still
an appreciable fraction.

The decay rate of \Eq{decay} is given by
\be
 \G_{H^+ \to S_i f \bar f'} = \frac{N_c}{\( 2 \pi\)^3} \frac{1}{32 M_{H^+}^3}
                 \int^{m_{12}^\text{2 max}}_{m_{12}^\text{2 min}} d m_{12}^2
                 \int^{m_{23}^\text{2 max}}_{m_{23}^\text{2 min}} d m_{23}^2 \, |\M_{H^+ \to S_i f \bar f'}|^2,
\label{Gff}
\ee
where $m_{12}$ and $m_{23}$ are kinematic variables defined as
\bea
 m_{12}^2 &=& M_{H^\pm}^2 + m_f^2 -2 M_{H^\pm} E_{f}, \\
 m_{23}^2 &=& M_{H^\pm}^2 + M_{S_i}^2 -2 M_{H^\pm} E_{S_i},
\eea
and the  integration limits are given by
\bea
 {m_{12}^\text{2 max}} &=& \( M_{H^\pm} - m_f \)^2, \\
 {m_{12}^\text{2 min}} &=& \( M_{S_i} + m_{f'} \)^2,
\eea
and
\bea
 {m_{23}^\text{2 max}} &=& \( E_2^*+E_3^* \)^2 - \(\sqrt{E_2^{*2}-m_{f'}^2} - \sqrt{E_3^{*2}-m_f^2} \)^2, \\
 {m_{23}^\text{2 min}} &=& \( E_2^*+E_3^* \)^2 - \(\sqrt{E_2^{*2}-m_{f'}^2} + \sqrt{E_3^{*2}-m_f^2} \)^2.
\eea
Here
\bea
 E_2^* &=& \frac{m_{12}^2 - M_{S_i}^2 + m_{f'}^2}{2 m_{12}}, \\
 E_3^* &=& \frac{M_{H^\pm}^2 - m_{12}^2 - m_f^2}{2 m_{12}},
\eea
and $m_f, m_{f'}$ are the outgoing fermion masses.
$N_c$ is the color number of the outgoing fermions.

The first integral in \Eq{Gff} is performed analytically, while the second one is performed numerically.
We give the exact result of \Eq{Gff} just in the limit $m_f, m_{f'} \ll \Delta M$ so that we can neglect fermionic masses,
\bea
&& \hspace{-1.1cm}
   \G_{H^+ \to S_i f \bar f'} \simeq
   \frac{N_c \eta_i^2 e^4}{24 s_W^4 M_W^2} \Bigg\{ {\D M} ({\D M}+2 M_{S_i})\times \nn\\
 &&
   \left[-2 \( \D M \)^2 ({\D M}+2 M_{S_i})^2+9 M_W^2 \left(({\D M}+M_{S_i})^2+M_{S_i}^2\right)-6    M_W^4\right]  \nn\\
   &&\left. \left. +3 M_W^2 \Bigg\{2 \left(({\D M}+M_{S_i})^2+M_{S_i}^2-M_W^2\right)   \text{X} \times \right. \right.  \nn\\
&& \left. \left. \left.
   \left[\tan ^{-1}\left(\frac{M_W^2-{\D M} ({\D M}+2  M_{S_i})}{\text{X}}\right)-
   \tan ^{-1}\left(\frac{M_W^2+{\D M} ({\D M}+2  M_{S_i})}{\text{X}}\right)\right] \right. \right. \right. \nn\\
&& -\left[\left(({\D M}+M_{S_i})^2-M_W^2 \right)^2+M_{S_i}^4-2 M_{S_i}^2 M_W^2\right] \log   \left(\frac{M_{S_i}^2}{({\D M}+M_{S_i})^2}
 \right)\Bigg\} \Bigg\}
\label{Gffexactm0}
\eea
where
\be
\text{X}=\sqrt{\left(M_W^2-\( \D M \)^2\right) \left(({\D M}+2  M_{S_i})^2-M_W^2\right)}
\ee
In order to have a better physical understanding of the process we consider the following limit
\be
 \d = \frac{\D M}{ M_{S_i}}  \ll 1
\ee
which is realized in case of displaced vertices. We expand \Eq{Gffexactm0} in powers of $\d$ and obtain an approximate expression
\be
 \G_{H^+ \to S_i f \bar f'} \simeq \frac{N_c \eta_i^2 e^4}{1920 \pi^3 s_w^4} \frac{ M_{S_i}^5 (2-3 \d) \d^5}{M_W^4} + {\cal O}(\d^7)
\label{Gffapprox}
\ee
\FIGURE[t!]{
\centering
\includegraphics[scale=0.35]{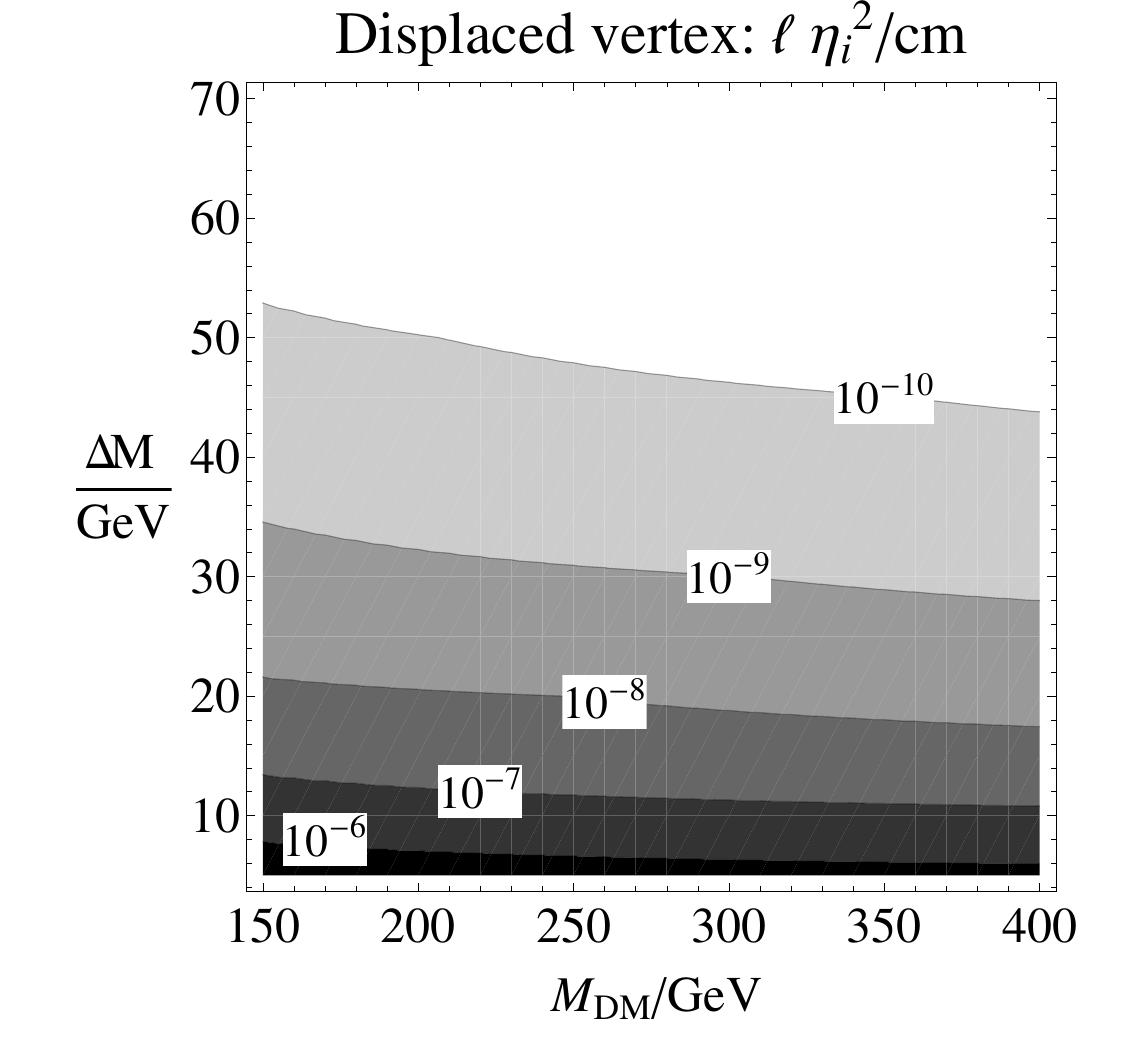}
\includegraphics[scale=0.35]{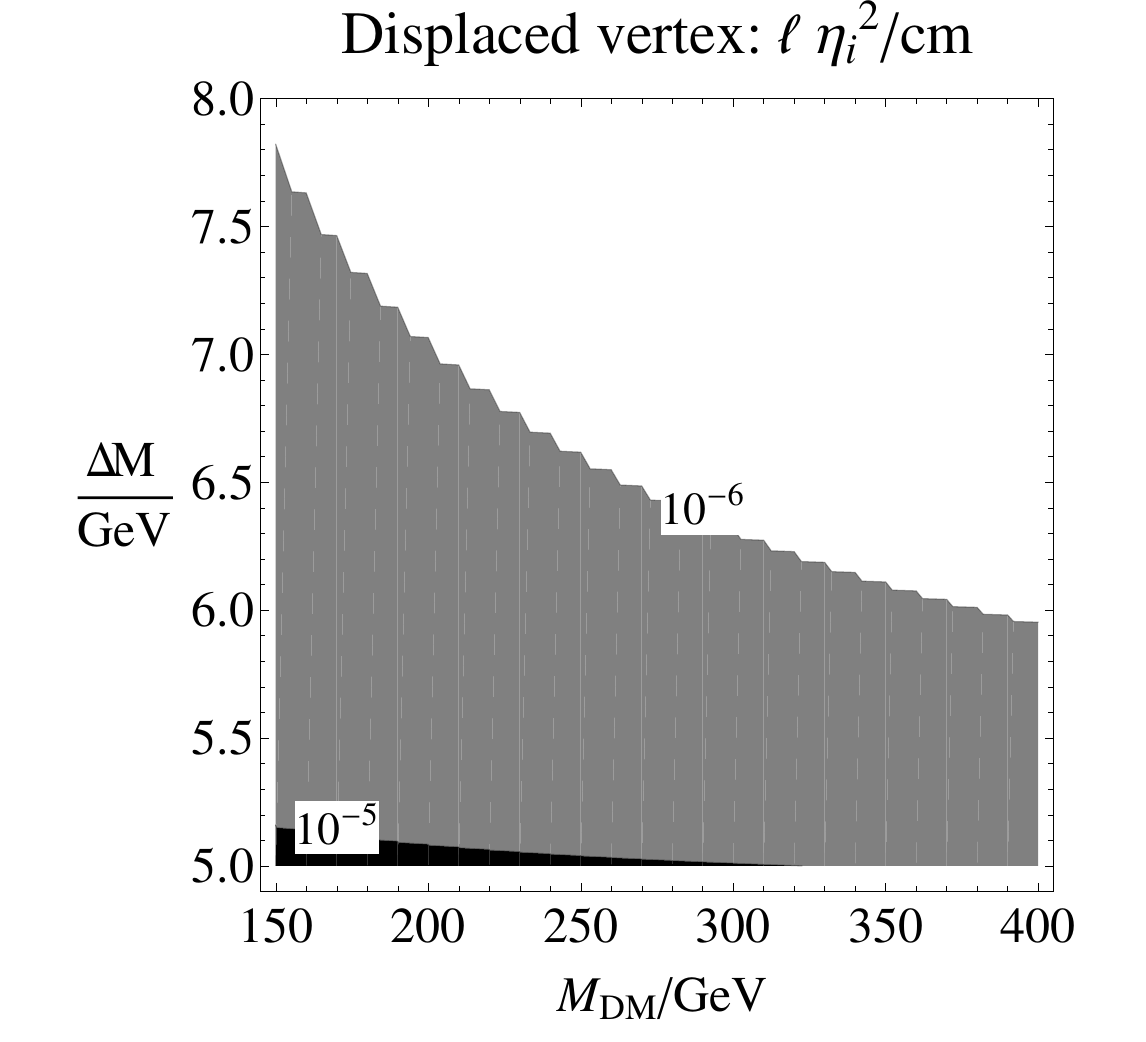}
\caption{The distance $\ell$ travelled by $H^\pm$ times $\eta_i^2$ for $E_{H^\pm}=1$ TeV
  as a  function of the DM mass $M_\text{DM}$ and the mass gap $\Delta M=M_{H^\pm}-M_\text{DM}$}
\label{Hpdisplacedgeneral}
}

The decay rate \Eq{Gffexactm0} is proportional to $\eta_i^2=s^2$
if the outgoing scalar is singlet-like (usually this is the lightest state) or to
$\eta_i^2=c^2$ if the outgoing scalar is doublet-like where
 $s (c)$ is the sine (cosine) of the mixing angle of the new scalar states.
The model dependence enters in the value of the $\eta_i$ parameter and in the number of $S_i$ possible states:
two for an inert doublet model \cite{Cao:2007rm,Dolle:2009ft}, four for our model.
From the approximated rate \Eq{Gffapprox} is clear the strong suppression induced by $\D M \ll M_{S_i}  $.
Moreover we notice that in the region $0< \d < 5/9$, and in particular in the region $\d \ll 1$, \Eq{Gffapprox} is a increasing function of
$\d$, in agreement with the kinematic behavior of \Eq{Gff} and \Eq{Gffexactm0}.

 Fig. \ref{Hpdisplacedgeneral}  shows model independent contour plots for
(the distance $\ell$)$\times$(the sine of singlet-doublet mixing angle squared, $ \eta_i^2$) travelled by $H^\pm$
as a function of the dark matter mass $M_\text{DM}$ and the mass gap $\Delta M=M_{H^\pm}-M_\text{DM}$.
In the left panel we consider large values for $\Delta M$ while in the right panel we assume
$\D M<8$ GeV.
One can see that, in order to get a macroscopic displaced vertex (for instance $\ell \gtrsim
0.1 $~mm), one needs both small values for
the mixing parameter $ \eta_i^2$ and the small mass gap. However, the latter needs not to be fine tuned to extreme values,
the mass gap of several GeV is quite natural.

\FIGURE[t!]{
\centering
\includegraphics[scale=0.3]{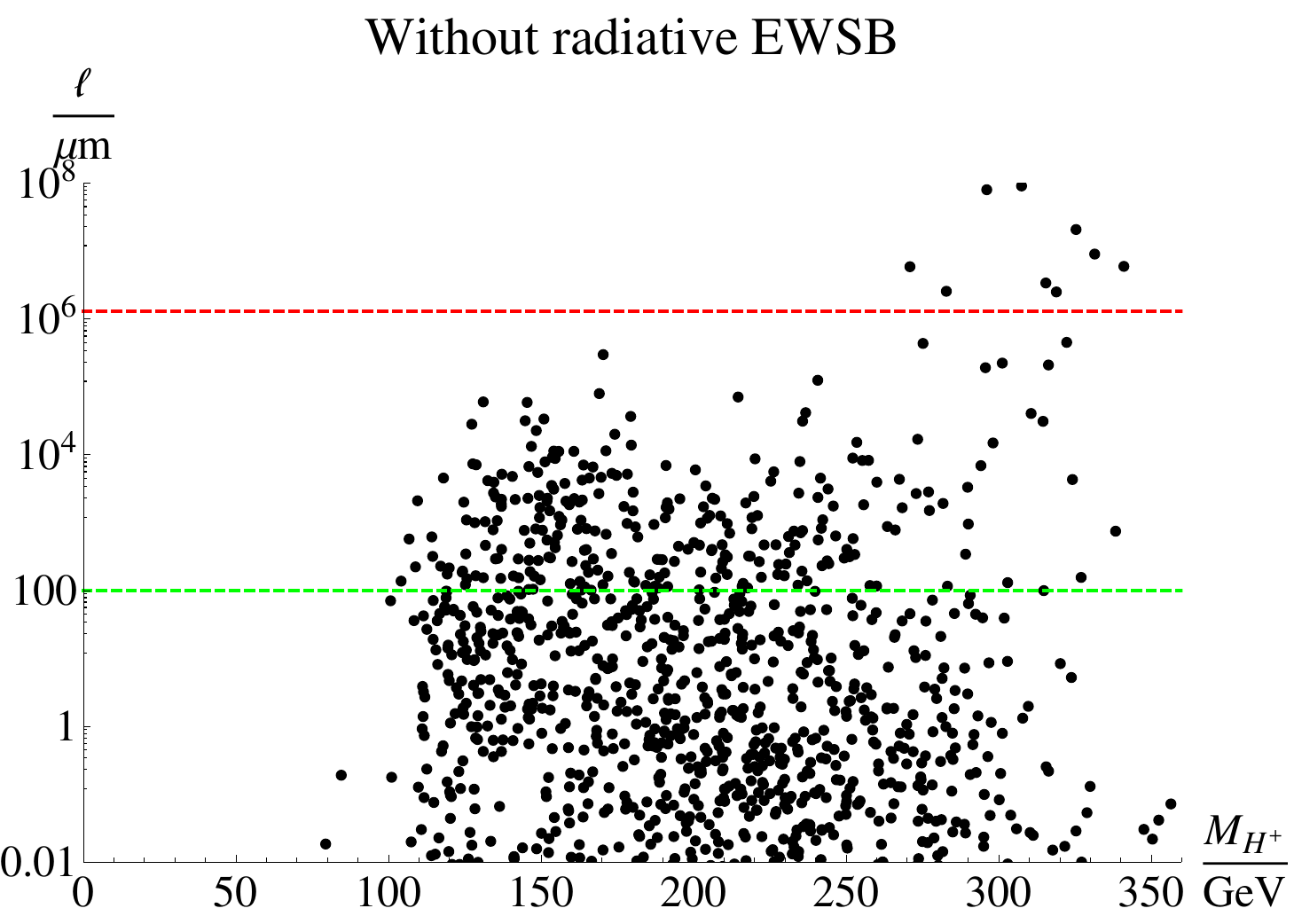}
\includegraphics[scale=0.3]{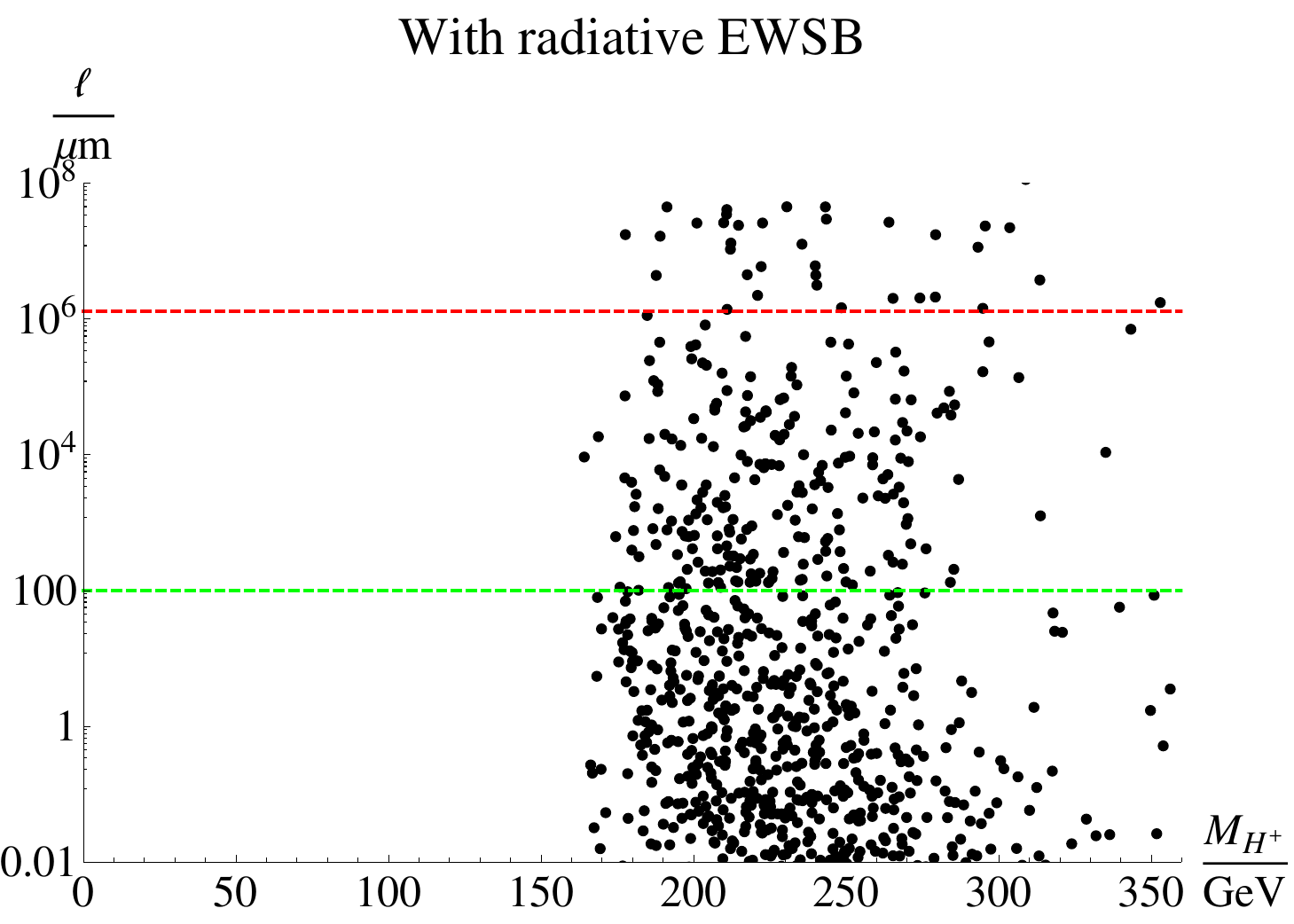}
\caption{Distance of  $H^\pm$ displaced vertex from the interaction point as a function of its
mass $M_{H^\pm}$ for $E_{H^\pm}=1$ TeV. In the left panel  radiative EWSB due to DM is not required
while in the right panel the radiative EWSB is required to occur. The region between dashed
lines is the CMS tracker radius.}
\label{Hpdisplaced}
}

\FIGURE[t!]{
\centering
\includegraphics[scale=0.3]{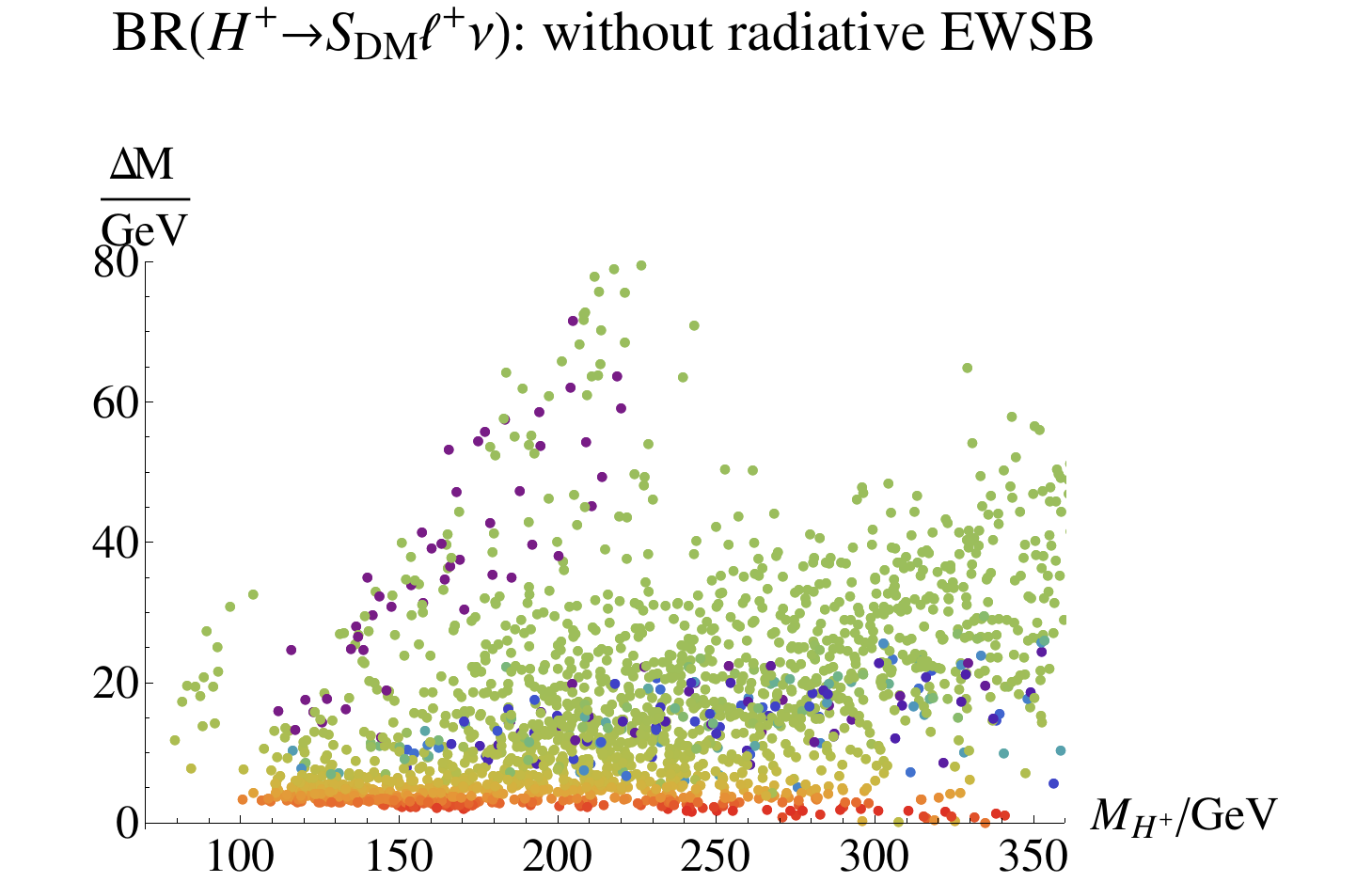}
\includegraphics[scale=0.3]{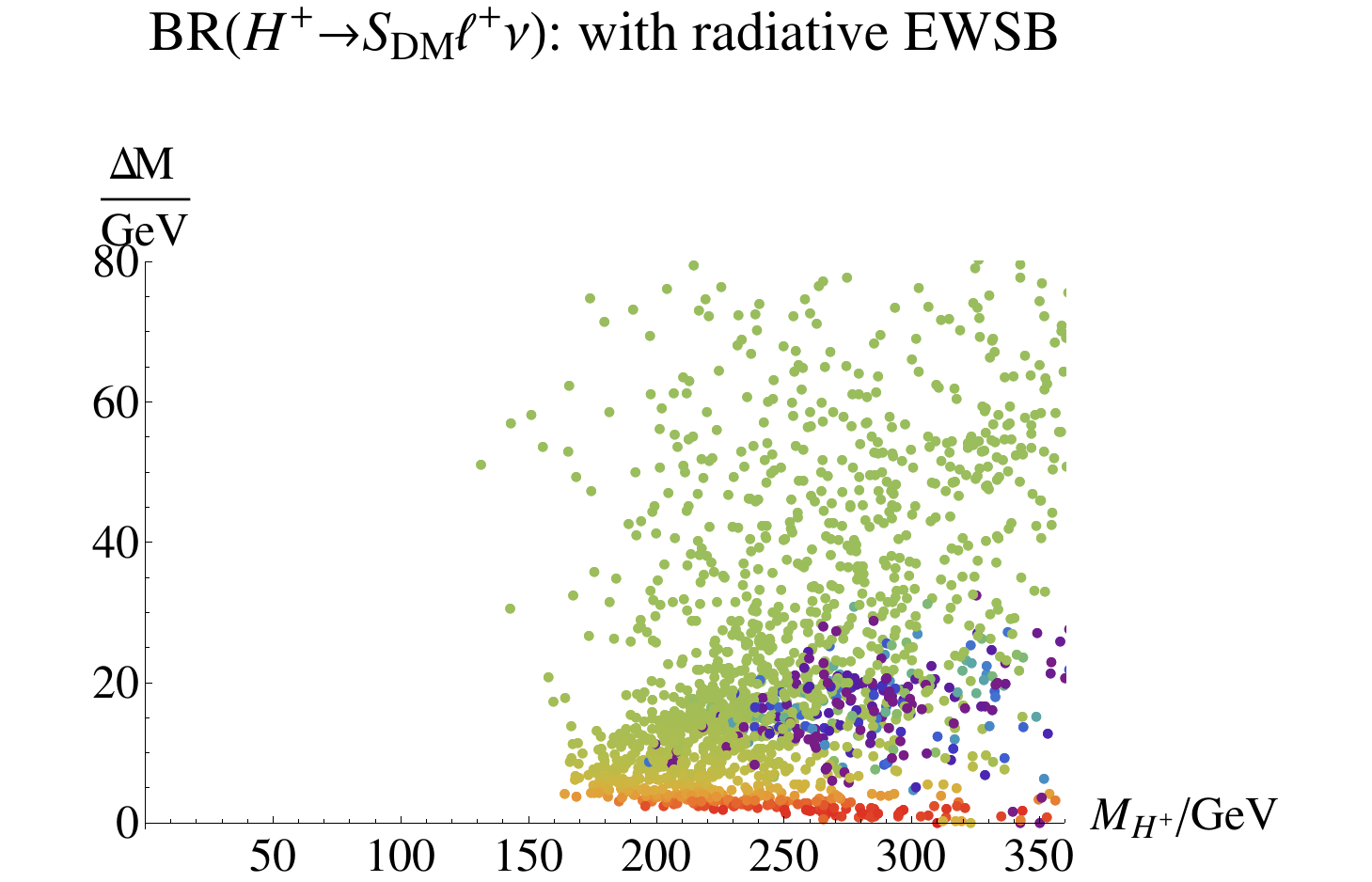}
\caption{Branching ratios for the $H^+ \to \sdm \ell^+ \n$ decay mode, where $\ell=e,\m$, without  (left panel) and with (right panel)
radiative EWSB. The branching ratios are represented by the colour code:  the red points mean BR~$\sim 0.2$, green points BR~$\sim 0.1$ and
 blue points BR~$\sim 0.02$. The branching ratios between those values are filled with the corresponding rainbow colours.}
\label{BRplotdisp}
}

Fig. \ref{Hpdisplaced} shows  the distance of displaced vertex in inert  $H^{\pm}$  decays  from the interaction point
for the same parameter points as in Fig. \ref{fig:cshp:lhc}.
Even if the distance is usually microscopic, for large region in the parameter space the charged Higgs displaced vertex can be
measured at the LHC experiments. In some cases $H^{\pm}$ are so long-lived that they may decay outside the detector.
In such cases DM is strongly singlet-like, that means $\eta  \to 0$ and there is an accidental  mass degeneracy
between $H^\pm$ and $S_{\text{DM}}$.
Those two experimental signatures are theoretically SM background-free and allow $H^\pm$ to be discovered at the LHC
up to the masses $M_{H^\pm}\lsim {\cal O}(300)$~GeV.
The latter estimate is based on our calculation of the production cross sections and decay branching ratios and
should be quantified with a detailed detector level simulation which is beyond the scope of this work.

If the only available decay modes of $H^\pm$ are $H^\pm \to S_{\text{DM}} f f' $ and $H^\pm \to S_{\text{NL}} f f' ,$  one has
$BR(H^\pm \to S_{\text{DM}} f f' )=BR(H^\pm \to S_{\text{NL}} f f' )$ and the branching ratios can be obtained just by counting
the available light SM quarks and leptons in the final state. However, if the decays to the heavier neutral scalar states $\snltw,\snlth$ become kinematically
available the branching ratios to different decay channels are highly model dependent. This happens when the mass splitting between the DM and
$H^\pm$ becomes large.

We plot in Fig. \ref{BRplotdisp}  the sum of branching ratios $BR(H^\pm \to S_{\text{DM}} \ell \nu)$  into the leptonic final states $\ell=e$ and $\ell=\mu$
as a function of $M_{H^{+}}$ and $\D M$. The branching ratios are expressed with the colour code as explained in the caption.
The red points confirm what we discussed in the previous paragraph:  for small mass splitting only the decays into $S_{\text{DM}}$
final state are possible and the branching ratio is essentially constant depending on the available SM fermions in the final state.
However, for large $\Delta M$ the branching ratio $BR(H^\pm \to S_{\text{DM}} \ell \nu)$ may vary in a considerably wide range.

  \TABLE[t]{
  \centering
 \begin{tabular}[h]{|c|c|c|c|c|c|c|c|c|c|c|}
  \hline
 & $M_{H^\pm}$  & $M_\text{DM}$   & $M_h$  & $\s^{qq}$ & $\s^{gg}$ &
$\lambda_3$ & $\eta$ & $\ell$ & R-EWSB \\
  \hline
 P1 &      80.99       & 60.75   & 166.9   & 276.1     & 890.5   &
$-0.21$ & 0.46 & 1.93$\times 10^{-6}$ & No \\

 P2 &       164.2   & 160.0  & 135.5 &  22.0  & 0.032     & $-0.114$
& 0.0074    & 9.22 & Yes   \\

 P3 &        196.6     & 192.1 & 140.8 &  11.3  & 0.032    & $-0.181$
& 3.9$\times 10^{-5}$ & 1.99$\times 10^5$  & Yes   \\
          \hline
 \end{tabular}
\vspace{-0.2cm}
  \caption{Benchmark points for the inert charged Higgs boson phenomenology at the LHC. The masses are given in units of GeV, the cross sections $\s^{qq,gg}$ for $pp \to H^+ H^-$
  in fb and the length $\ell$ in mm.}
}

\section{Benchmark points for the constrained scalar DM model}

To study the charged inert Higgs boson pair production at the LHC  we propose three benchmark points with a distinctive phenomenology.
The points are summarized in Table~1. In all points the charged Higgs is light, the pair production cross section is
large and the correct amount of predominantly singlet DM is produced in thermal freeze-out at early Universe.

The main feature of  P1 is that $H^\pm$ mass is below the radiative EWSB threshold. If
such a light  $H^\pm$ is discovered, the
EW symmetry must be broken explicitly.
In this point the pair production cross section is dominated by the sub-process
 \rfn{eq:ggHH} which is proportional to the single parameter $\lambda_3.$ Because  the cross section of  \rfn{eq:qqHH}
 depends only on the $H^\pm$ mass, its contribution to the total cross section can be computed accurately.
 Therefore, in the case of P1  the measurement of  $H^\pm$ pair production cross section at the LHC
 implies a measurement of $\lambda_3$ with a high accuracy. Because in this point EWSB occurs radiatively due to RG effects,
 the measurement of $\lambda_3$, together with the mass determination via \Eq{mch},  offers a consistency check of the model.

The second feature of P1 is that $H^\pm$ decay promptly and there is no displaced vertex. Nevertheless this point $H^\pm$ can be
discovered over the huge SM $W^\pm$ background using two additional experimental signatures.
First, the subsequent decays of the $H^\pm$ decay products,  $S_{\text{NL}}\to S_{\text{DM}} f\bar f,$
necessarily produce additional displaced vertices that allow to reconstruct the full $H^\pm$ decay chain.
Second, the SM Higgs boson total width as well as the branching fractions are modified due to the existence
of the SM Higgs decays to the $H^+H^-$ final states with $BR(H^+ H^-)=0.21.$ In the P1 the SM Higgs boson width is 0.39~GeV while the
SM prediction is 0.31~GeV.

The point P2 is characteristic to our model as  $H^\pm$ decay inside the tracker of the LHC experiment
leaving displaced vertices.
The experimental signature of P2 is that the charged
track of $H^\pm$  breaks into a  charged lepton track and missing energy. This is theoretically a background-free signature.
In this case the EWSB occurs radiatively while the SM Higgs boson physics is well described by the SM predictions.

In the case of P3 the inert $H^\pm$ is so long-lived that it crosses the tracker of the LHC experiment completely and decays outside the detector.
This case can be discovered by a slow charged track of $H^\pm .$ Thus the phenomenology of P3 resembles the phenomenology of charged
R-hadron \cite{Ball:2007zza} rather that the charged Higgs boson.  However, its production cross section is determined by weak interactions not
by the strong interaction as in the case of a typical R-hadron.

\section{Conclusions}

We have studied the inert charged Higgs boson production at the LHC in the context of constrained
scalar DM model. The previous similar works, performed in the
context of inert doublet model, have shown that the  $H^+H^-$ final states cannot be seen at the LHC over the huge $W^+W^-$
background. However, the inert doublet model is just one particular limit of a general matter-parity induced
scalar DM scenario. In the constrained scalar DM model studied in this work the lightest DM scalar
is predicted to be predominantly singlet. Due to the small singlet-doublet mixing the $H^\pm$ lifetime can be macroscopically long
 if, in addition, the charged Higgs and the DM particle have a small mass difference.

We have  recomputed the $H^+H^-,$ $H^\pm S_i$ production cross sections in $pp$ collisions at the LHC. We have included
a gluon-gluon one-loop contribution depicted in  Fig.~\ref{CHprod} and shown that it may give the dominant contribution to
 the total cross section for light $H^\pm$ due to the SM Higgs boson resonances.
We have required the production of correct amount of DM of the Universe in thermal freeze-out and  analyzed this scenario in two
distinctive cases: when the
EWSB occurs radiatively due to the DM interactions with the SM Higgs boson and when EW symmetry is broken explicitly.
We show that in an appreciable  part of the parameter space $H^\pm$ can be long-lived and decay via \rfn{decay} at a macroscopic
distance from the interaction point. This signature is theoretically background-free and allows $H^+H^-$ to be discovered at the LHC
up to the masses $M_{H^\pm}\lsim {\cal O} (300)$~GeV.
To test this scenario we have proposed three benchmark points with different charged Higgs phenomenology.
If, however, $H^\pm$ are short-lived, their decays to the next-to-lightest neutral scalar, namely $H^\pm \to S_{\text{NL}} f f' ,$
will be followed by the decays $S_{\text{NL}}\to S_{\text{DM}} f\bar f.$
In the latter case the experimental signatures of the $H^+H^-$ pair production include {\it two} displaced
$\ell^+\ell^-$ or $jj$ vertices.
Those unique experimental signatures allow one to discover  the inert charged Higgs
over the SM background at the LHC.

The numerical estimates in this paper are based on our theoretical calculations of the production cross sections
and decay branching ratios and should be quantified with a detailed detector level simulation which is beyond the scope of this work.

\vspace{0.5cm}

\acknowledgments{
This work was supported by the following grants: ESF 8090, JD164 and
SF0690030s09.
KH is grateful for the support by the Academy of Finland (Project No. 115032).
}

\end{document}